**Title:**
Electrical transport and percolation in magnetoresistive manganite / insulating oxide composites: case of $La_{0.7}Ca_{0.3}MnO_3$ / $Mn_3O_4$

**Authors:**
B. Vertruyen[*], R. Cloots, M. Ausloos, J.-F. Fagnard, Ph. Vanderbemden
SUPRATECS, University of Liege, Sart-Tilman, B-4000 Liege, Belgium
* Corresponding author:
e-mail: b.vertruyen@ulg.ac.be, phone : + 32 4 3663452, fax : + 32 4 3663413

**Abstract**
We report the results of electrical resistivity measurements carried out on well-sintered $La_{0.7}Ca_{0.3}MnO_3$ / $Mn_3O_4$ composite samples with almost constant composition of the magnetoresistive manganite phase ($La_{0.7}Ca_{0.3}MnO_3$). A percolation threshold ($\Phi_c$) occurs when the $La_{0.7}Ca_{0.3}MnO_3$ volume fraction is ~ 0.19. The dependence of the electrical resistivity $\rho$ as a function of $La_{0.7}Ca_{0.3}MnO_3$ volume fraction ($f_{LCMO}$) can be described by percolation-like phenomenological equations. Fitting the conducting regime ($f_{LCMO} > \Phi_c$) by the percolation power law $\rho \propto (f_{LCMO} - \Phi_C)^{-t}$ returns a critical exponent $t$ value of 2.0±0.2 at room temperature and 2.6±0.2 at 5 K. The increase of $t$ is ascribed to the influence of the grain boundaries on the electrical conduction process at low temperature.

**Text**

The colossal magnetoresistance (CMR)[1] in $La_{1-y}A_yMnO_{3-d}$ compounds (A=Ca,Sr) is strongly influenced by the microstructure.[1-3] The presence of grain boundaries in polycrystalline materials leads to a significant magnetoresistance at low fields and all temperatures below the paramagnetic/ferromagnetic transition temperature $T_C$.[4] Several authors have tried to increase the low-field magnetoresistance (LFMR) by mixing a CMR manganite phase with an insulating oxide secondary phase.[5-16] When manganite-insulator composites are sintered at high temperature to achieve densification, ionic diffusion usually results in a pronounced shift of the manganite composition.[5-10] As a result, it is not possible to analyze the electrical properties as a function of the volume percentage of each phase, since the composition of each phase varies significantly throughout the series of samples. In order to prevent the interdiffusion phenomena, some authors have prepared composites using very short heat treatments,[12-16] so that the composition of the manganite phase was not affected. However the short sintering treatments result in large porosity of the samples and poor connectivity between the manganite grains; it is difficult to take these effects properly into account when parameterizing the system for a description of the electrical conductivity behavior.

In a previous work,[17] we have shown that well-sintered $La_{0.7}Ca_{0.3}MnO_3$/$Mn_3O_4$ composite samples can be obtained with almost constant manganite composition throughout the series. A one-step spray drying synthesis[17,18] followed by a long thermal treatment at 1300°C results in the formation of dense samples containing $Mn_3O_4$ and a LCMO manganite phase. The stoichiometry of the $La_{0.7}Ca_{0.3}MnO_3$ (LCMO) phase is not significantly affected by the $Mn_3O_4$ secondary phase, as proved[17] by comparing the properties of the two end members of the composite series (cell parameters, Curie temperature and saturation magnetization).

In the present work, the dependence of the electrical resistivity properties as a function of the $La_{0.7}Ca_{0.3}MnO_3$ volume fraction is analyzed in the framework of percolation theory. The microstructure of the samples was characterized by scanning electron microscopy (Philips FEG ESEM XL-30) of polished cross-sections. The volume resistivity of the high-resistivity samples was measured with a Keithley 617 Electrometer across the thickness of disk-shaped pellets using silver paste contacts attached to the circular faces. Surface leakage currents were found to be negligible with respect to volume current, whence no guard ring was used. The resistance of the low-resistivity samples was measured with a HP 34420 nanovoltmeter using the conventional four-point technique on thin bar-shaped samples cut from the pellets using a wire saw. In the intermediate resistivity regime ($10^2 < \rho < 10^6$ Ω cm), both geometries were used and results were compared to each other. Electrical resistance *vs.* temperature $R(T)$ curves were recorded in a Quantum Design PPMS.

Scanning electron micrographs (Figure 1) revealed three types of microstructures as the $Mn_3O_4$ content increases: (i) $Mn_3O_4$ islands in a LCMO matrix, (ii) a labyrinthic pattern of the two phases and (iii) LCMO islands in a $Mn_3O_4$ matrix. In $Mn_3O_4$-rich samples, a significant porosity was observed. Density measurements by Archimedes' method[21] were carried out in order to calculate the volume fractions of each phase, i.e. the LCMO phase, the $Mn_3O_4$ phase and the total (open+closed) porosity. The dimensionless volume fractions are defined in such a way that $f_{LCMO} + f_{Mn3O4} + f_{porosity} = 1$. In LCMO-rich samples ($f_{LCMO} > 0.3$), $f_{porosity}$ does not exceed 0.10 whereas in $Mn_3O_4$-rich samples ($f_{LCMO} < 0.3$), $f_{porosity}$ lies in the 0.20-0.25 range. A systematic over-estimation of the absolute $f_{LCMO}$ values cannot be excluded with this method[19] and is estimated to be smaller than 0.03. The important point is the repeatability of the $f_{LCMO}$ determination, which was within the $[f \pm 0.002]$ interval.

The electrical resistance of the composite materials depends on the sample porosity and on the respective volume fractions and resistivities of the LCMO and $Mn_3O_4$ phases. Resistivity values of pure LCMO samples prepared using the synthesis technique described above are typically on the order of $10^{-2}$ Ω cm (for 5 K < $T$ < 300 K), whereas pure $Mn_3O_4$ samples display electrical resistivity values on the order of $10^8$ Ω cm at room temperature and higher values for $T$ < 300 K. Air pores are insulating. The composite materials can thus be considered as made up of a high-conductivity part (LCMO) and a low-conductivity part ($Mn_3O_4$ + pores). The results will be discussed in terms of the conducting phase volume fraction ($f_{LCMO}$) as the unique parameter characterizing the different composite materials.

The temperature dependences of the zero-field resistivity of samples with $f_{LCMO}$ = 0.18, 0.20, 0.40, 0.80 and 0.92 are shown in Figure 2 with the resistivity axis in logarithmic scale. All samples with $f_{LCMO} > \sim 0.19$ present a resistivity transition from a semiconducting-like behavior above the transition peak temperature $T_p$ to a metallic-like behavior below $T_p$. The $T_p$ value of the sample with $f_{LCMO} \sim 0.20$ is only ~ 20 K lower than the value for the manganite-pure sample. This is *much* smaller than the corresponding $T_p$ drops observed in other insulator-manganite composite series densified at high temperature.[7,9,10,12] The difference between $T_p$ and the Curie temperature $T_C$ (determined from magnetization measurements)[17] never exceeds 5 K, as usually observed in well-sintered $La_{0.7}Ca_{0.3}MnO_3$.[1,2] When $f_{LCMO}$ decreases from 0.92 towards 0.20, a low-temperature bump progressively grows and eventually becomes higher than the $T_p$ maximum in the case of the sample with $f_{LCMO} \sim$ 0.20 (see inset of Figure 2). Such a behavior is usually considered as a grain-boundary-related contribution in manganite polycrystalline materials.[2,20] On the contrary, the samples with $f_{LCMO} < \sim 0.19$ do not display any resistive transition in the temperature range where the electrical resistance could be accurately measured with the electrometer (R < 100 GΩ).

The resistivity data of the samples in the semiconducting-like state can be fitted by a simple law $\rho = \rho_\infty \exp(E_0/kT)$.[1] The $E_0$ values for the samples presenting a resistive transition are in the 90-140 meV range. For the high-resistivity samples with $f_{LCMO} < \sim 0.19$, $E_0$ increases to values lying between 400 and 500 meV.

Figure 3 shows the resistivity of the composite samples at 300 K as a function of the LCMO volume fraction $f_{LCMO}$. The resistivity axis is in logarithmic scale. A sharp increase of four orders of magnitude in resistivity occurs at $f_{LCMO} \sim 0.19$ (dashed line) and clearly establishes the existence of a percolation threshold. The resistivity vs. $f_{LCMO}$ data at T = 5 K is shown in the inset of Figure 3. For $f_{LCMO} > \sim 0.19$, the behavior is similar to the 300 K data, with a sharp resistivity increase occurring at the same percolation threshold $f_{LCMO} \sim 0.19$. The resistivity values for $f_{LCMO} < \sim 0.19$ were too high to be measured accurately at low temperatures. Experimentally, resistivity values determined on samples with almost identical $f_{LCMO}$ were found to agree within a factor two. Such variations are observed even in manganite-pure samples and probably result from a small error in the geometrical factor estimation (mainly due to the finite size of voltage contacts) and the imperfect reproducibility of the pressing and sintering steps. In the immediate vicinity of the percolation threshold [$f_{LCMO} \sim (0.18 \pm 0.01)$], there is also a size effect of the contacts. In the case of the pellet measurement, the contact area is ~ 10 times larger and the distance between contacts is ~ 5 times smaller than in the case of bar-shaped samples. Therefore the probability of finding a conducting filamentary path is significantly larger in the case of the pellet configuration. This may result in overall resistivity values several orders of magnitude smaller for the pellet than for the bar-shaped samples. Bar-shaped configuration was found to lead to much better reproducibility than the pellet configuration and the resistivity data shown in Figure 3 for these samples correspond to the average of four-point resistivity measurements performed on three bars cut from each pellet.

The 0.19 experimental percolation threshold found for our LCMO/$Mn_3O_4$ composites is similar to the values calculated for regular periodic lattices: it was first observed by Scher and Zallen[21] that calculations for all usual 3D lattice types predict critical volume fractions $\Phi_c$ in the range 0.16±0.02. However theoretical predictions for regular lattices do not necessarily apply to continuum media and we should also compare the 0.19 percolation threshold to literature data for other manganite-insulator composite systems with no or little interdiffusion between the phases. Balcells et al.[13] report a percolation threshold $f_{LSMO} = 0.20$ for a $La_{2/3}Sr_{1/3}MnO_3/CeO_2$ composite series, in good agreement with the value obtained here. However, most papers report experimental percolation thresholds far from the Scher and Zallen range: $f_{LCMO}$ between 0.92 and 0.85 for $La_{0.67}Ca_{0.33}MnO_3/Al_2O_3$,[14,16] $f_{LCMO} \sim 0.60$ for $La_{0.67}Ca_{0.33}MnO_3/SrTiO_3$ [12] and $f_{LSMO} \sim 0.50$-0.55 for $La_{0.7}Sr_{0.3}MnO_3$/borosilicate glass.[15] Several reasons may explain the dispersion of these experimental values. First, the significant porosity is sometimes not taken into account when calculating the volume fraction. Second, the poor connectivity between neighboring manganite grains is likely to affect the $\Phi_c$ values since some of the neighboring conducting grains fail to make contact.[22] Third, the significant difference in grain size between the conducting and the insulating phases in references[12,15] results in a highly non-uniform distribution of the two phases in the system. Such mechanisms can yield critical volume fractions far from the 0.16 theoretical prediction.[23]

In view of the above, the rather good agreement between the experimental percolation threshold of our composites and the Scher and Zallen theoretical range for lattice percolation suggests that the one-step spray-drying synthesis leads to more or less random "auto-organization" of the two phases in the samples. Our experimental percolation threshold is also an indirect indication of the good connectivity between the LCMO grains in the composites. This is further supported by the very small temperature difference (< 5 K) between the electrical transition temperature $T_p$ and the Curie temperature $T_C$ in the studied samples. For these reasons, the $La_{0.7}Ca_{0.3}MnO_3/Mn_3O_4$ system appears as a good benchmark to study the applicability of percolation laws to manganite-insulator composite systems.

The high value (~$10^{10}$ at T = 300 K) of the resistivity ratio between the "insulating" ($Mn_3O_4$) and the "conducting" (LCMO) phases allows us to attempt to describe the electrical resistivity $\rho$ of the composites by percolation laws:[24]
In the conducting regime ($f_{LCMO} > \Phi_c$):
$$\rho = \rho_1 \left( f_{LCMO} - \Phi_C \right)^{-t} \qquad (1)$$
In the insulating regime ($f_{LCMO} < \Phi_c$):
$$\rho = \rho_2 \left( \Phi_C - f_{LCMO} \right)^{s} \qquad (2)$$
where $\Phi_c$ is the critical volume fraction of the conducting phase, $t$ and $s$ are critical exponents and $\rho_1$ and $\rho_2$ are adjustable parameters.

We have used equations (1) and (2) to fit the resistivity data in the range $f_{LCMO} > 0.19$ and $f_{LCMO} < 0.19$ respectively. The best fits are shown as plain lines in Figure 3 (300 K) and inset (5 K). The experimental trend in the whole $f_{LCMO}$ range is reasonably described by percolation-like laws with $\Phi_c = 0.191$. Although percolation theory is supposed to apply to the vicinity of the percolation threshold, it is in fact frequent that percolation laws can model data obtained far from the percolation threshold.[23] The fitting returns critical exponent $t$ values of 2.0 at 300 K and 2.6 at 5 K. The error on $t$ was estimated by observing that a 0.2 variation of $t$ decreases markedly the fit quality. The fit of the high-resistivity regime at 300 K yields a critical exponent $s$ of 1.4, but this value cannot be compared to results for biphasic insulator-conductor composites because the "high resistivity phase" is actually made up of a mixture of $Mn_3O_4$ and porosity: when $f_{LCMO} < \Phi_c$, charge carriers travel both in LCMO and $Mn_3O_4$ but not in air pores and it would be necessary to consider three phases.

In three-dimensional continuum composites, the conductivity exponent $t$ is often found in the 1.6-2.0 range[23,24] but larger values of $t$ are also frequently reported,[24] so the $t$ values found here at 300 K (*i.e.* 2.0±0.2) and 5 K (*i.e.* 2.6±0.2) are not abnormally high. The perceptible difference between the two $t$ values may result from the contribution of the grain boundaries to the resistivity at low temperatures, as explained hereafter. For $f_{LCMO} > \Phi_c$, the electrical current flows through a continuous LCMO path by crossing the grain boundaries between neighboring LCMO grains. Above $T_C$ (*e.g.* at 300 K), the LCMO phase is paramagnetic and Mn spins are disordered, both in the grains and at the grain boundaries. Below $T_C$ (e.g. at 5 K), the LCMO phase is ferromagnetic and the main contribution to the electrical resistance results from the grain boundaries, where the parallel alignment of the Mn spins is disrupted.[1,2] In the present work, the low temperature bump in the $\rho(T)$ curve becomes more pronounced when $f_{LCMO}$ approaches $\Phi_c$, indicating an increasing contribution of the grain boundaries[2,20] to the conduction process as $f_{LCMO}$ is reduced from 1 to 0.19. As a result, below $T_C$, the "effective resistivity" of the LCMO phase cannot be considered as constant in the whole $f_{LCMO} > \Phi_c$ range. This increase of the LCMO effective resistivity reasonably explains the observed increase of the conductivity exponent $t$ at 5 K with respect to the 300 K value.

BV thanks the F.N.R.S. in Belgium for a Postdoctoral researcher fellowship. Part of this work was supported by the European Network of Excellence FAME.

**Figure 1**: Electron micrographs of polished cross-sections of samples with decreasing LCMO volume fractions : (a) 0.83 (b) 0.40 (c) 0.18. Light grey and dark grey regions correspond to LCMO and $Mn_3O_4$, respectively.

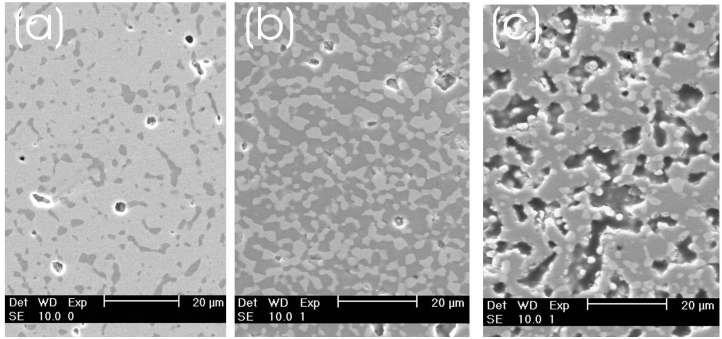

**Figure 2**: Main graph: Temperature dependence of the zero-field resistivity of samples containing manganite volume fractions $f_{LCMO}$ ranging from 0.18 to 0.92. $T_p$ of each sample is indicated by an arrow. Inset: Same data presented in linear scale for $f_{LCMO} \sim 0.20$.

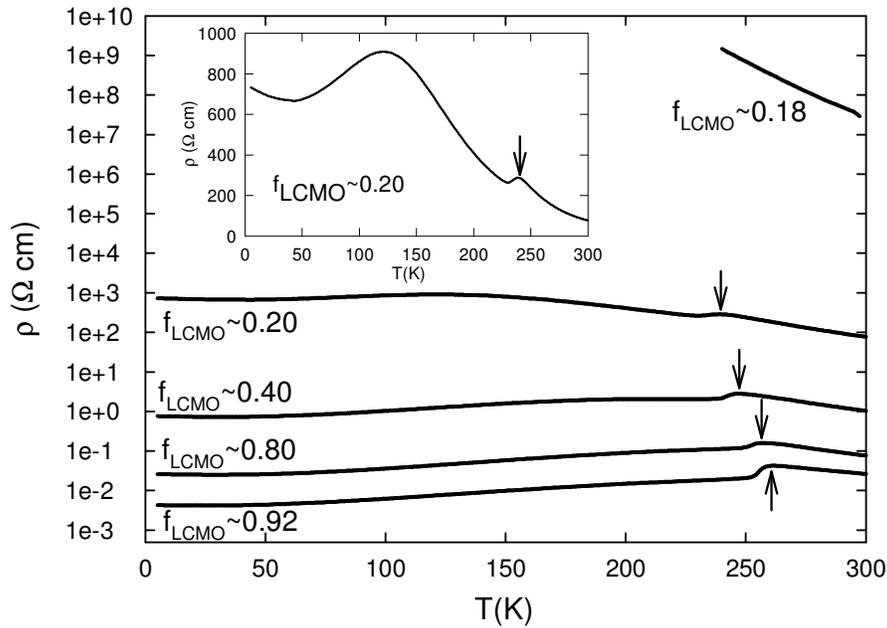

**Figure 3**: Electrical resistivity at 300 K as a function of the LCMO volume fraction $f_{LCMO}$. Inset : Electrical resistivity at 5 K for $f_{LCMO} < \sim 0.19$. The plain lines correspond to fits by different equations (see text for details).

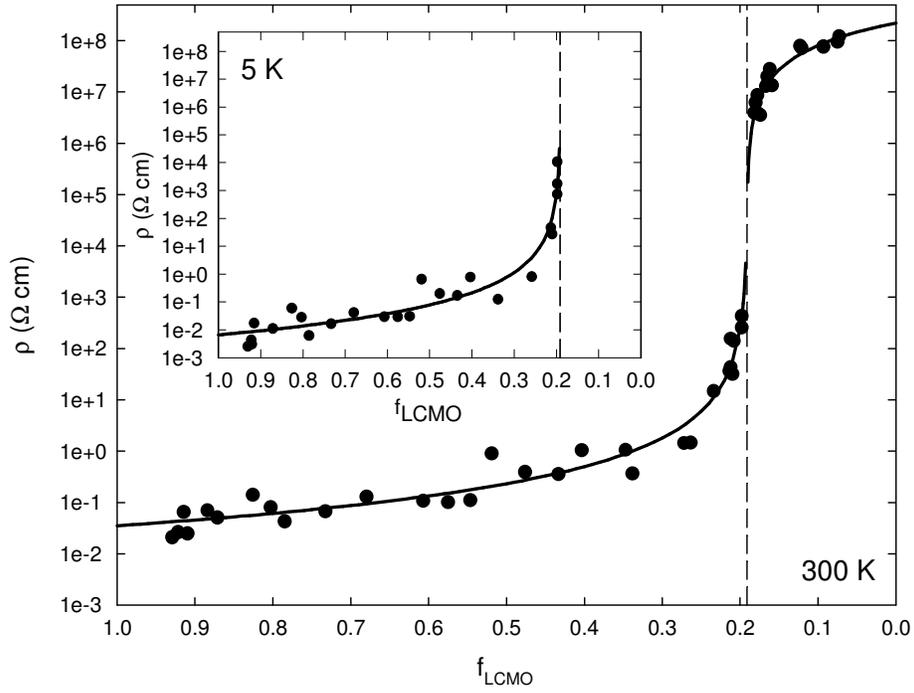